\documentclass[a4paper,11pt]{article}

\usepackage{pos}
\usepackage{lipsum} 
\usepackage{wrapfig}
\title{Sensitivity of IceCube-Gen2 to measure flavor composition of Astrophysical neutrinos}

\ShortTitle{Sensitivity of IceCube-Gen2 to measure Flavor composition of neutrinos}

\author{The IceCube-Gen2 Collaboration \\{\normalsize \normalfont(a complete list of authors can be found at the end of the proceedings)}\\}

\emailAdd{neha.lad@desy.de}

\abstract{

The observation of an astrophysical neutrino flux in IceCube and its detection capability to separate between the different neutrino flavors has led IceCube to constraint the flavor content of this flux. IceCube-Gen2 is the planned extension of the current IceCube detector, which will be about 8 times larger than the current instrumented volume. In this work, we study the sensitivity of IceCube-Gen2 to the astrophysical neutrino flavor composition and investigate its tau neutrino identification capabilities. We apply the IceCube analysis on a simulated IceCube-Gen2 dataset that mimics the High Energy Starting Event (HESE) classification. Reconstructions are performed using sensors that have 3 times higher quantum efficiency and isotropic angular acceptance compared to the current IceCube optical modules. We present the projected sensitivity for 10 years of data on constraining the flavor ratio of the astrophysical neutrino flux at Earth by IceCube-Gen2.

\vspace{4mm}
{\bfseries Corresponding authors:}
Neha Lad$^{1,2,*}$\\
{$^{1}$ \itshape Deutsches Elektronen-Synchrotron DESY: Zeuthen, Germany}\\
{$^{2}$ \itshape Institut für Physik, Humboldt-Universität zu Berlin: Berlin, Germany}\\[4mm]
$^*$ Presenter

\ConferenceLogo{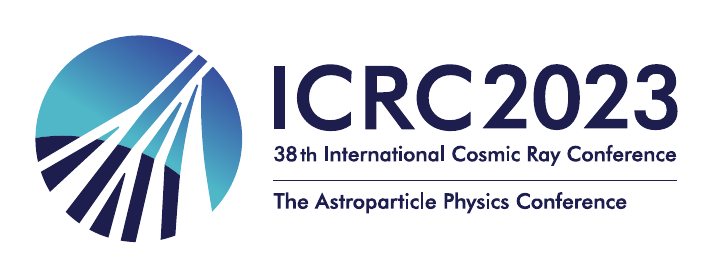}

\FullConference{The 38th International Cosmic Ray Conference (ICRC2023)\\ 26 July -- 3 August, 2023\\ Nagoya, Japan}
}

\begin{document}

\maketitle

\section{Introduction to IceCube-Gen2}\label{Intro}
IceCube is a cubic-kilometer neutrino detector buried deep in the ice of the geographic South Pole \cite{Aartsen:2016nxy}. The first ever discovery IceCube reported came within only 3 years of its active time, when it discovered diffuse astrophysical neutrino flux measured on earth\cite{discovery_paper}. By utilizing the special properties of neutrinos, IceCube offers a unique window into understanding the fundamental workings of our universe.
IceCube-Gen2 \cite{whitepaper} will be an extension of the existing IceCube array designed to investigate the high-energy neutrino sky. With help of 120 new strings, 240m apart from each other, 9600 new optical modules will be deployed around current IceCube, making its volume about eight times bigger. Additionally, a large surface array \cite{SurfaceArray} will be implemented to improve the identification of neutrino interactions in the Southern sky by enhancing the veto of cosmic-ray (CR) air showers and lowering the energy threshold. The primary objective of the radio array \cite{whitepaper} is to detect neutrinos with energies surpassing 100 PeV. 

\section{Flavor Composition of Astrophysical Neutrinos}\label{Flavor}
While discovery of astrophysical neutrinos provides a valuable insight into diffuse spectrum of neutrinos, it does not offer direct insights into the sources or the processes through which they were produced. Depending on mechanism at production site ($\nu_e:\nu_{\mu}:\nu_{\tau}$), measured fractions of  neutrino flavors on earth is different due to neutrino oscillations. Thus, studying the flavor content of these neutrinos can significantly deepen our understanding of the diverse high-energy particle-generating phenomena in the universe.  

IceCube, by using 7.5 years of High Energy Starting Events (HESE) \cite{HESE7}, for the first time put constraints on the flavor content of astrophysical neutrinos with non-zero $\nu_{\tau}$ fraction \cite{Juliana}. The HESE sample is an all-sky, all-flavor neutrino sample with very high astrophysical purity. In this study, IceCube’s ability to identify the three different event topologies, cascades, tracks and double cascades was taken advantage of, as a particle identifier. All of these topologies are specific to the flavor of neutrino and its interaction type. Charged-current (CC) $\nu_e$ interactions and all-flavor Neutral-current (NC) interactions produce single cascades. Tracks on the other hand are produced by CC $\nu_{\mu}$ and atmospheric muons (excluding those muons originating CC $\nu_{\tau}$). Most interesting and difficult to reconstruct topology is double cascade, a topology specific to only CC $\nu_{\tau}$ events, where a $\nu_{\tau}$ interacts with nucleon to produce a tau lepton, which then travels some distance in ice ($\propto E_\tau$) and decays to produce charged leptons or multiple hadrons. Excluding the muonic decay of tau, all other decay channels produces a second hadronic or electromagnetic cascade, this yields a topology with two causally-connected cascades that can be reconstructed using a multi-source likelihood method. 

It this proceeding, we discuss IceCube-Gen2's capability to detect $\nu_{\tau}$ induced double cascades and its sensitivity to measure the flavor composition of the astrophysical neutrino spectrum. Moreover, we also show its sensitivity to detect a changing flavor composition as a function of neutrino energy, which will allow us to distinguish different acceleration scenarios and source environments expected within GRBs, AGN cores, or AGN jets \cite{flavortheory}.

\section{Simulation and Software}\label{Simulation}
\begin{wrapfigure}{l}{0.5\textwidth}
  \begin{center}
    \includegraphics[width=0.3\textwidth]{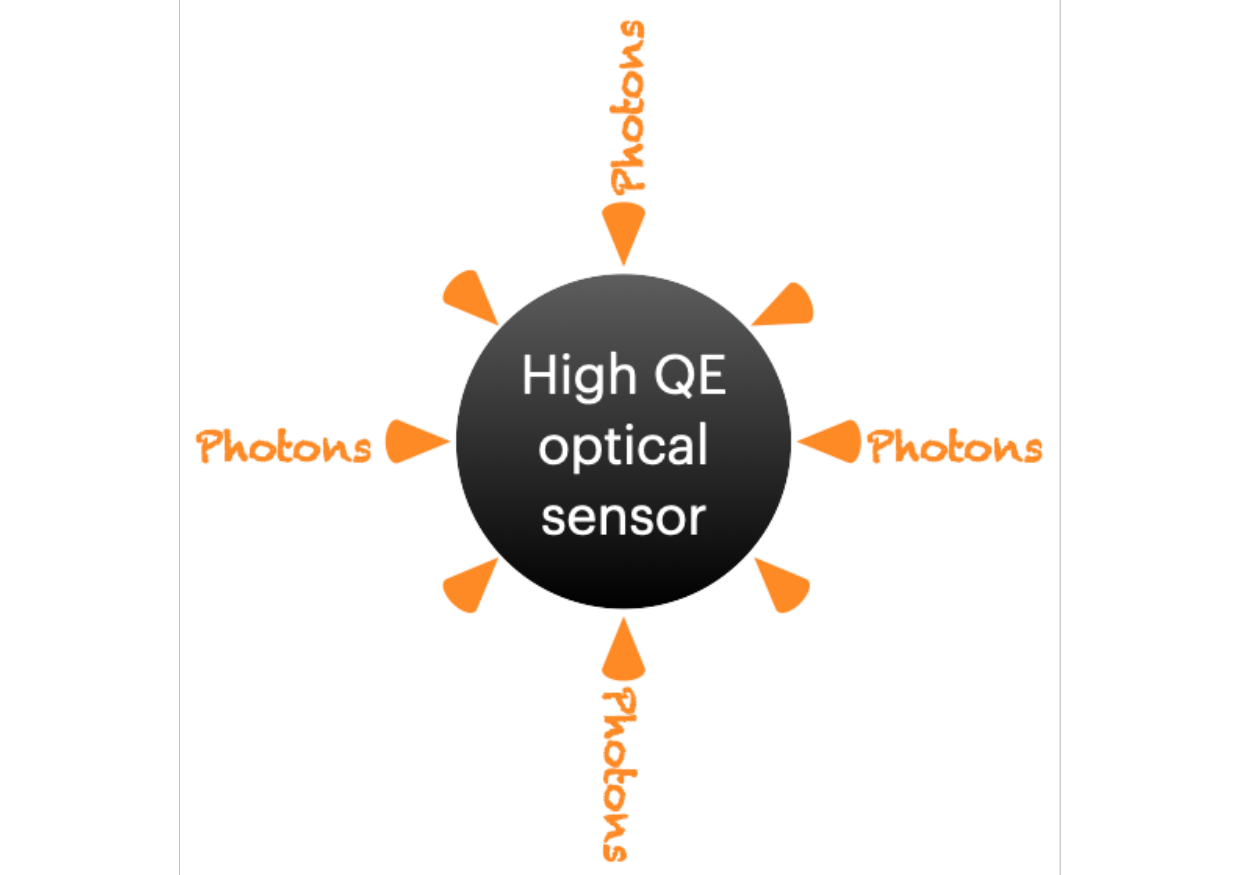}
  \end{center}
  \caption{Conceptual representation of Simulated sensor with isotropic angular acceptance (iso-pDOM)}
  \label{fig:isopdom}
\end{wrapfigure}

IceCube-Gen2 will incorporate innovative optical sensors like mDOMs and D-Eggs \cite{whitepaper}. The multi-PMT DOM (mDOM) is an optical module equipped with 24 3-inch photo-multiplayers (PMTs). This particular type of DOM offers a photocathode area that is about 2 times larger than that of IceCube pDOM. It provides isotropic sensitivity and enables the extraction of information from individual "pixels" (24 PMTs). To facilitate the utilization of established techniques for reconstructing simulated events with the new sensors, for this study, mDOMs were simulated as pDOM-type modules. These simulated modules feature approximately three times higher quantum efficiency and isotropic angular acceptance (Fig.~\ref{fig:isopdom}). Chosen Geometry of the detector strings for this study is "sunflower" geometry, with strings 240m apart from eachother, a most optimal geometry for IceCube-Gen2 \cite{Anastasiia}. Monte Carlo events were produced assuming these so called iso-pDOM for all-flavor neutrinos having energy between 100TeV to 50PeV.

To apply a similar particle identifier that was used in HESE-7.5 to reconstruct events, a HESE-like event selection was implemented. We only select events that have reconstructed energy greater than 100 TeV and has neutrino interaction vertex starting inside the detector. Chain of reconstruction and sub-classifications remains identical to that used before. 

Sensitivity projections for flavor measurement were performed using \texttt{toise} \cite{toise}, a framework designed to perform sensitivity studies for neutrino detectors. It uses parameterised detector responses which can be further used to project sensitivities such as effective area, energy and angular resolutions etc.

\section{Sensitivity to measure flavor}\label{Sensitivity}
\begin{figure}[h!]

    \centering
    \includegraphics[width=0.55\columnwidth]{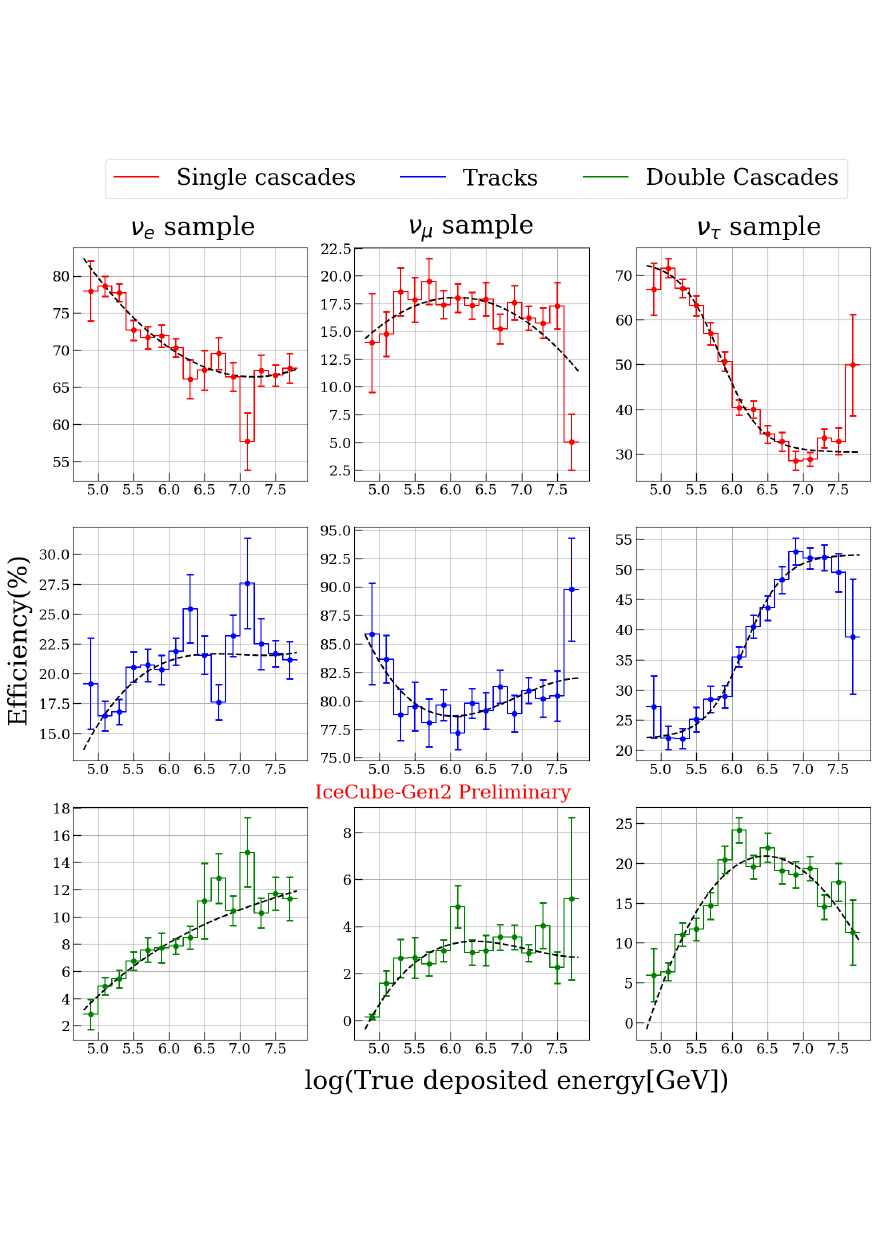}
    \caption{Classification Efficiency : Three subplot columns are true neutrino flavors, where each energy bin (true monte-carlo energy) contains fraction of topologies, summing to a 100\%. Diagonal plots shows flavor identification efficiency of the classifier whereas off-diagonal plots shows mis-identification fractions.}
\label{fig:classeff}
\end{figure}

\begin{figure}[h!]
  \centering
    \includegraphics[width=0.6\columnwidth]{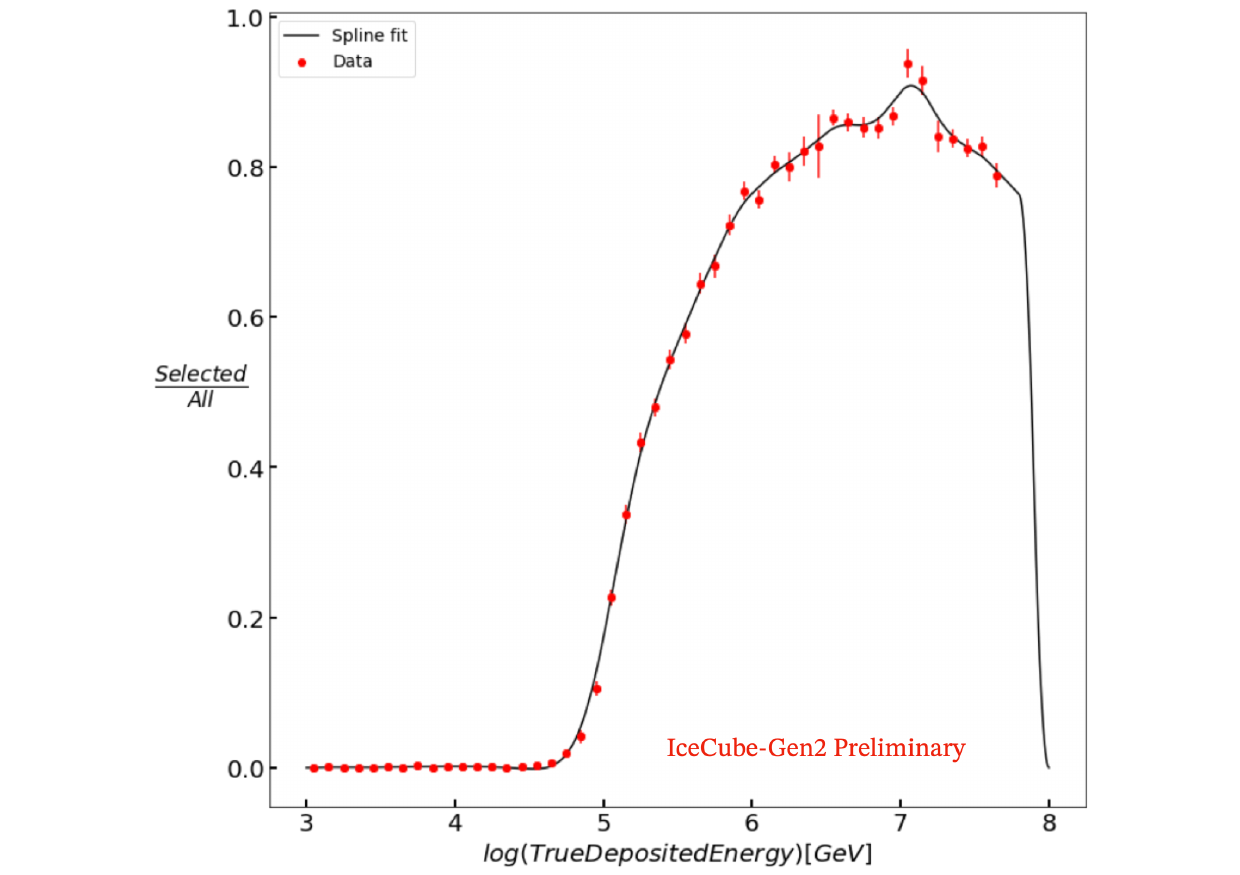}
    \caption{Selection Efficiency : Ratio of neutrinos that got classified into a topology to all the neutrinos that interacted in active volume. Data here refers to monte-carlo events per energy bin.}
\label{fig:selecteff}
\end{figure}

\texttt{toise}, as mentioned before, requires parameterized detector responses to project sensitivities. For flavor measurement, it requires classification efficiency (Fig.~\ref{fig:classeff}) and selection efficiency(Fig.~\ref{fig:selecteff}), which are parameterized and used by \texttt{toise} to create asimov sensitivity contours. Classification efficiency is defined as the fraction of topology per energy bin for a given flavor of neutrino. Diagonal plots in Fig.~\ref{fig:classeff} defines flavor identification efficiency and off-diagonal plots represent mis-identification fractions. Selection efficiency, on the other hand, is defined as, the ratio of the number of neutrinos that end up being classified into any topology to number of neutrinos that interacted in active volume. 

\begin{figure}[h!]
  \centering
    \includegraphics[width=0.6\columnwidth]{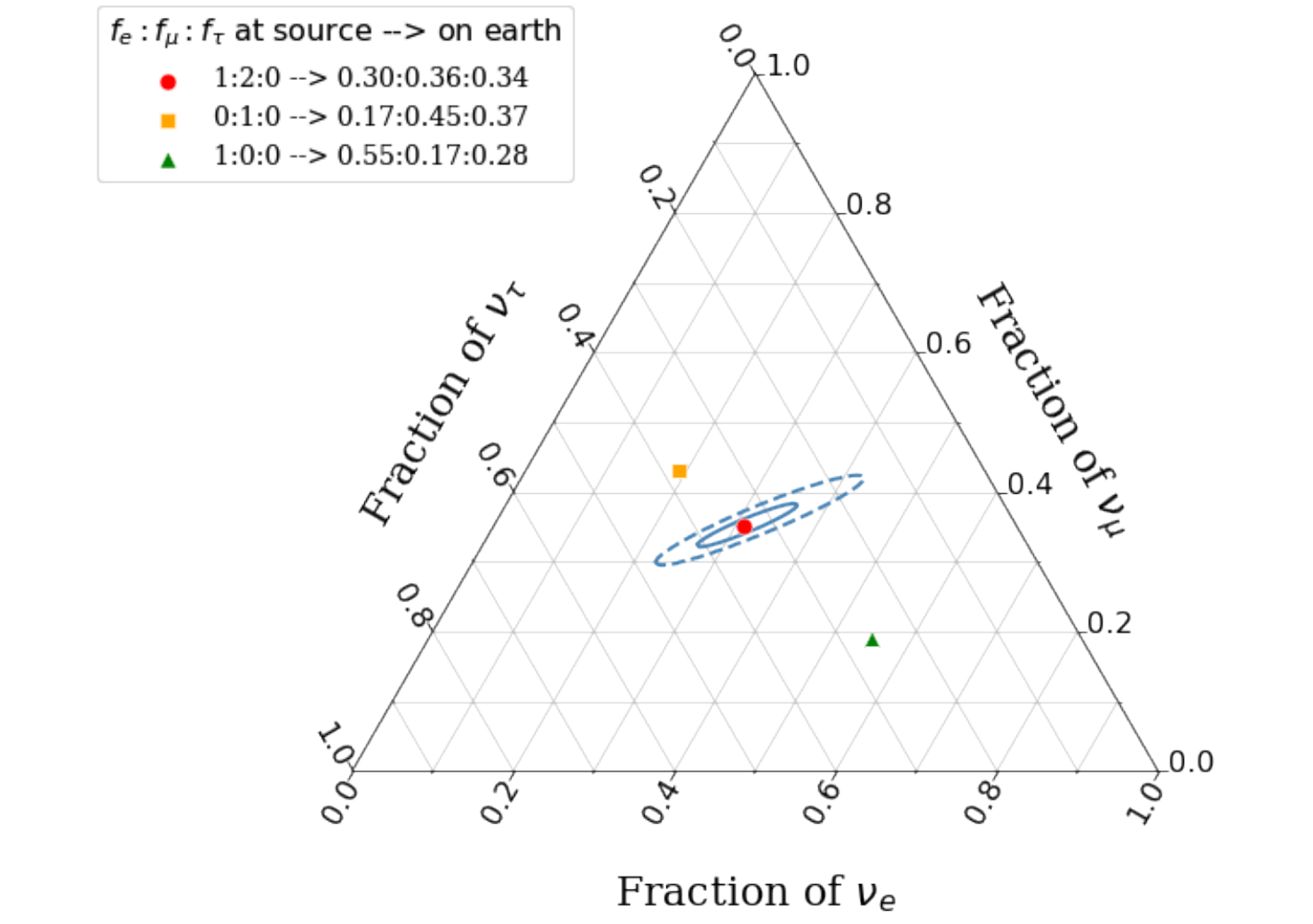}
    \caption{Projected sensitivity of IceCube-Gen2 to measure flavor composition of Astrophysical neutrino with 10 years of its lifetime. The dashed (solid) outlines depict the corresponding 99\% (68\%) constraints. Diffuse neutrino spectrum is assumed to be single power law with index of -2.5 and per-flavor norm of 2.3 \cite{IceCube:2015gsk}.}
    \label{fig:Flavortriangle}
\end{figure}

\begin{figure}[h!]
  \centering
    \includegraphics[width=0.6\columnwidth]{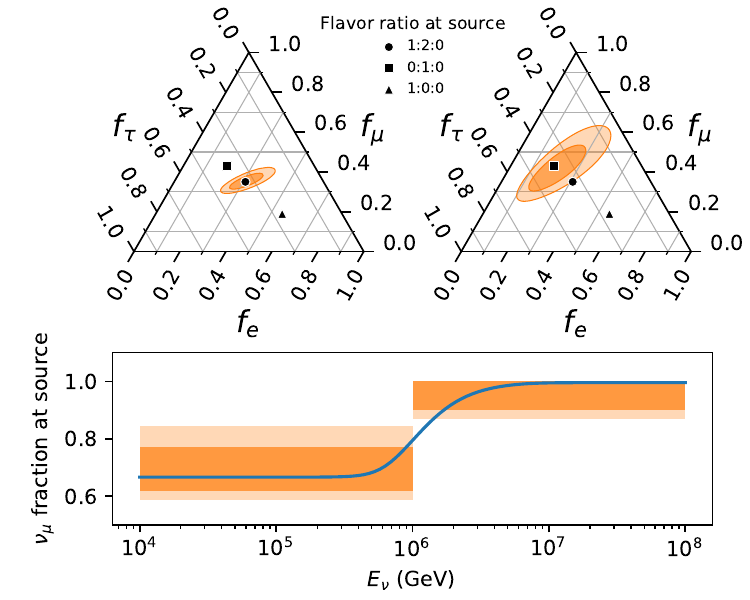}
    \caption{The bottom section displays the proportion of $\nu_{\mu}$ at the source based on energy, with the assumption that the muon critical energy is 2 PeV. The error bars represent the 68\% confidence level limitations on the $\nu_{\mu}$ fraction below and above 1 PeV, derived from the observed flavor composition of $\nu_{\mu}$ at Earth using IceCube-Gen2 and assuming standard oscillations. In the upper sections, the dark (light) shaded regions depict the corresponding 68\% (99\%) constraints, without making any assumptions about the mixing matrix.}
    \label{fig:muondamped}
\end{figure}

Result of this study is shown in Fig.~\ref{fig:Flavortriangle} and Fig.~\ref{fig:muondamped}. Fig.~\ref{fig:Flavortriangle} shows flavor measurement sensitivity of IceCube-Gen2 assuming the pion-decay scenario of neutrino production at high energy astrophysical sources. Above a critical energy, the flux of electron neutrinos is suppressed due to the presence of sufficiently strong magnetic fields, resulting in the damping of muons. Fig.~\ref{fig:muondamped} shows sensitivity of IceCube-Gen2 to measure flavor ratios of neutrinos on earth assuming muon-damped scenario of neutrino production at the sources. It is to be noted that sensitivity projected here assumes a 'globalfit', that is, a combined fit of starting events and thoroughgoing tracks \cite{10diffuse}, a concept which is currently being realised in IceCube to measure spectrum  \cite{GlobalFit:2023icrc} and flavor composition \cite{Tauicrc2023} of diffuse astrophysical neutrinos. 

\section{Conclusions and Outlook}\label{Conclusion}
IceCube-Gen2 is a planned extension of current IceCube neutrino telescope at south pole. With about 8 times larger detection volume, multi PMT optical modules and extended radio detector arrays, IceCube-Gen2 will be sensitive to even EeV neutrinos. The rate of observable $\nu_{\tau}$ events in the IceCube detector is limited by the density of instrumentation because the distance between two cascades is typically much smaller than the distance between two DOMs. By significantly increasing its size, IceCube-Gen2 will be capable of observing a considerably higher rate of these so called 'double cascade' events, resulting in more stringent constraints on the flavor ratio.

In this study, we use monte carlo simulations, which were generated using a sensor that best represents mDOM like sensor (isotropic light acceptance and high quantum efficiency). Using parameterised detector responses such as classification and selection efficiency of particle identifier in \texttt{toise}, a frame work designed to project sensitivities of future neutrino detectors, one can derive how sensitive IceCube-Gen2 will be to measure the flavor ratio of astrophysical neutrinos with its 10 years of livetime. By utilizing the substantial samples of neutrinos encompassing all flavors in IceCube-Gen2, we will be able to examine the energy-dependent variations in the flavor ratio across an extensive range of energies. The heightened sensitivity in detecting alterations in the composition of flavors with respect to neutrino energy will enable us to differentiate between various acceleration scenarios and source environments that are anticipated within various high energy particle accelerators of the cosmos.
 
\bibliographystyle{ICRC}
\bibliography{references}

%

\clearpage

\section*{Full Author List: IceCube-Gen2 Collaboration}

\scriptsize
\noindent
R. Abbasi$^{17}$,
M. Ackermann$^{76}$,
J. Adams$^{22}$,
S. K. Agarwalla$^{47,\: 77}$,
J. A. Aguilar$^{12}$,
M. Ahlers$^{26}$,
J.M. Alameddine$^{27}$,
N. M. Amin$^{53}$,
K. Andeen$^{50}$,
G. Anton$^{30}$
% ...
\\
\\
% ...
$^{12}$ Universit{\'e} Libre de Bruxelles, Science Faculty CP230, B-1050 Brussels, Belgium \\
% ...
$^{17}$ Department of Physics, Loyola University Chicago, Chicago, IL 60660, USA \\
% ...
$^{22}$ Dept. of Physics and Astronomy, University of Canterbury, Private Bag 4800, Christchurch, New Zealand \\
% ...
$^{26}$ Niels Bohr Institute, University of Copenhagen, DK-2100 Copenhagen, Denmark \\
$^{27}$ Dept. of Physics, TU Dortmund University, D-44221 Dortmund, Germany \\
% ...
$^{30}$ Erlangen Centre for Astroparticle Physics, Friedrich-Alexander-Universit{\"a}t Erlangen-N{\"u}rnberg, D-91058 Erlangen, Germany \\
% ...
$^{47}$ Dept. of Physics and Wisconsin IceCube Particle Astrophysics Center, University of Wisconsin{\textendash}Madison, Madison, WI 53706, USA \\
% ...
$^{50}$ Department of Physics, Marquette University, Milwaukee, WI, 53201, USA \\
% ...
$^{53}$ Bartol Research Institute and Dept. of Physics and Astronomy, University of Delaware, Newark, DE 19716, USA \\
% ...
$^{76}$ Deutsches Elektronen-Synchrotron DESY, Platanenallee 6, 15738 Zeuthen, Germany  \\
$^{77}$ Institute of Physics, Sachivalaya Marg, Sainik School Post, Bhubaneswar 751005, India \\
% ...

\subsection*{Acknowledgements}

\noindent
The authors gratefully acknowledge the support from the following agencies and institutions:
USA {\textendash} U.S. National Science Foundation-Office of Polar Programs,
U.S. National Science Foundation-Physics Division,
U.S. National Science Foundation-EPSCoR,
Wisconsin Alumni Research Foundation,
Center for High Throughput Computing (CHTC) at the University of Wisconsin{\textendash}Madison,
Open Science Grid (OSG),
Advanced Cyberinfrastructure Coordination Ecosystem: Services {\&} Support (ACCESS),
Frontera computing project at the Texas Advanced Computing Center,
U.S. Department of Energy-National Energy Research Scientific Computing Center,
Particle astrophysics research computing center at the University of Maryland,
Institute for Cyber-Enabled Research at Michigan State University,
and Astroparticle physics computational facility at Marquette University;
Belgium {\textendash} Funds for Scientific Research (FRS-FNRS and FWO),
FWO Odysseus and Big Science programmes,
and Belgian Federal Science Policy Office (Belspo);
Germany {\textendash} Bundesministerium f{\"u}r Bildung und Forschung (BMBF),
Deutsche Forschungsgemeinschaft (DFG),
Helmholtz Alliance for Astroparticle Physics (HAP),
Initiative and Networking Fund of the Helmholtz Association,
Deutsches Elektronen Synchrotron (DESY),
and High Performance Computing cluster of the RWTH Aachen;
Sweden {\textendash} Swedish Research Council,
Swedish Polar Research Secretariat,
Swedish National Infrastructure for Computing (SNIC),
and Knut and Alice Wallenberg Foundation;
European Union {\textendash} EGI Advanced Computing for research;
Australia {\textendash} Australian Research Council;
Canada {\textendash} Natural Sciences and Engineering Research Council of Canada,
Calcul Qu{\'e}bec, Compute Ontario, Canada Foundation for Innovation, WestGrid, and Compute Canada;
Denmark {\textendash} Villum Fonden, Carlsberg Foundation, and European Commission;
New Zealand {\textendash} Marsden Fund;
Japan {\textendash} Japan Society for Promotion of Science (JSPS)
and Institute for Global Prominent Research (IGPR) of Chiba University;
Korea {\textendash} National Research Foundation of Korea (NRF);
Switzerland {\textendash} Swiss National Science Foundation (SNSF);
United Kingdom {\textendash} Department of Physics, University of Oxford.

\end{document}